\documentclass[useAMS,usenatbib]{mn2e}
\usepackage{graphicx}
\usepackage{ulem}
\usepackage{amsmath}

\title[Formation of S-type planets in close binaries:scattering induced tidal capture]{Formation of S-type planets in close binaries: scattering induced tidal capture of circumbinary planets}
\author[Yan-Xiang Gong \& Jianghui Ji]{Yan-Xiang Gong$^{1,2}$\thanks{yxgong@pmo.ac.cn}, Jianghui Ji$^{1}$\thanks{jijh@pmo.ac.cn}\\
$^{1}$CAS Key Laboratory of Planetary Sciences, Purple Mountain Observatory, Chinese Academy of Sciences, Nanjing 210008, China\\
$^{2}$College of Physics and Electronic Engineering, Taishan University, Taian 271000, China}
\begin{document}

\date{Received 2017 July 26; in original form 2017 October 30}

\pagerange{\pageref{firstpage}--\pageref{lastpage}} \pubyear{2018}

\maketitle

\label{firstpage}

\begin{abstract}
Although several S-type and P-type planets in binary systems were discovered in past years, S-type planets have not yet been found in close binaries with an orbital separation not more than 5 au. Recent studies suggest that S-type planets in close binaries may be detected through high-accuracy observations. However, nowadays planet formation theories imply that it is difficult for S-type planets in close binaries systems to form in situ. In this work, we extensively perform numerical simulations to explore scenarios of planet-planet scattering among circumbinary planets and subsequent tidal capture in various binary configurations, to examine whether the mechanism can play a part in producing such kind of planets. Our results show that this mechanism is robust. The maximum capture probability is $\sim 10\%$, which can be comparable to the tidal capture probability of hot Jupiters in single star systems. The capture probability is related to binary configurations, where a smaller eccentricity or a low mass ratio of the binary will lead to a larger probability of capture, and vice versa. Furthermore, we find that S-type planets with retrograde orbits can be naturally produced via capture process. These planets on retrograde orbits can help us distinguish in situ formation and post-capture origin for S-type planet in close binaries systems. The forthcoming missions (PLATO) will provide the opportunity and feasibility to detect such planets. Our work provides several suggestions for selecting target binaries in search for S-type planets in the near future.

\end{abstract}


\section{Introduction}

To date, more than 120 exoplanets have been found in binary star systems (\textit{http://www.univie.ac.at/adg/schwarz/ multiple.html}). According to the orbital configuration of binary-planet system, planets in binaries can be divided into two categories \footnote{A third type refers to planets that orbit near the $L_{4}$ or $L_{5}$ triangular Lagrangian points of binary. In this case, the mass ratio of the binary $\mu  = {{{M_2}} \mathord{\left/
 {\vphantom {{{M_2}} {\left( {{M_1} + {M_2}} \right)}}} \right.
 \kern-\nulldelimiterspace} {\left( {{M_1} + {M_2}} \right)}}$ must be less than 0.04 for motion about these points to be linearly stable.}: S-type (Satellite-like orbit), in which the planet's orbit encircles either of the stars of a binary, whereas the other is P-type (Planetary orbits), which the planet revolves around double stars \citep{Dvorak1986}. P-type planets are referred to circumbinary planets (CBP).

For S-type planets, it is generally believed that the companion star in a wide binary ($\rho > 100$ au) has little influence on their formation. However, it is not the case for close binaries. By exploring 382 Kepler Objects of Interest (KOIs), \citet{Kraus2016} revealed the planet occurrence rate in close binaries with a separation $<$ 47 au is only 0.34 times that of wider binaries or single stars. This indicates that the close binary companions have ruinous influence on the formation of S-type planets. Close binaries play a major role in the formation of planets, which has been investigated by many theoretical analyses (see \citet{Thebault2015} and references therein). Despite numerous theoretical hurdles to their formation, several known close binaries such as $\gamma$ Cep \citep{Hatzes2003} and Kepler-420 (AB) \citep{Santerne2014} are found clearly to host S-type planets. Statistical analysis of KOIs further showed that some planets may form in binary systems with projected separations as tight as $\rho = 2-3$ au \citep{Kraus2016}. Planets that have formed in such dynamically active environments provide crucial constraints for theories of binary and planet formation. However, there are still many unsolved mysteries for such objects.

Figure 1 shows the orbital period versus mass ratio $q_{B}=M_{2}/M_{1}$ distribution of the binaries harboring exoplanets.
The red circles represent S-type planets, whereas the blue circles denote P-type planets. The size of each circle is plotted to be proportional to ${m_{p}}^{1/3}$ of the planet, with respect to Jupiter's mass. As seen in Figure 1, S-type planets have not yet been found in the binaries with a period $P<1000$ days. Kepler-420 (AB) has a minimum Semimajor Axis (SMA) of 5.3 au (corresponding to an orbital period of 3430 days) in which only one S-type planet has been discovered \citep{Santerne2014}. However, over 3,000 eclipsing binaries with $P<1000$ days have been found by \textit{Kepler} telescope \citep{Kirk2016}, where Figure 1 exhibits their period vs. number distribution.

\begin{figure*}
\includegraphics[scale=0.7]{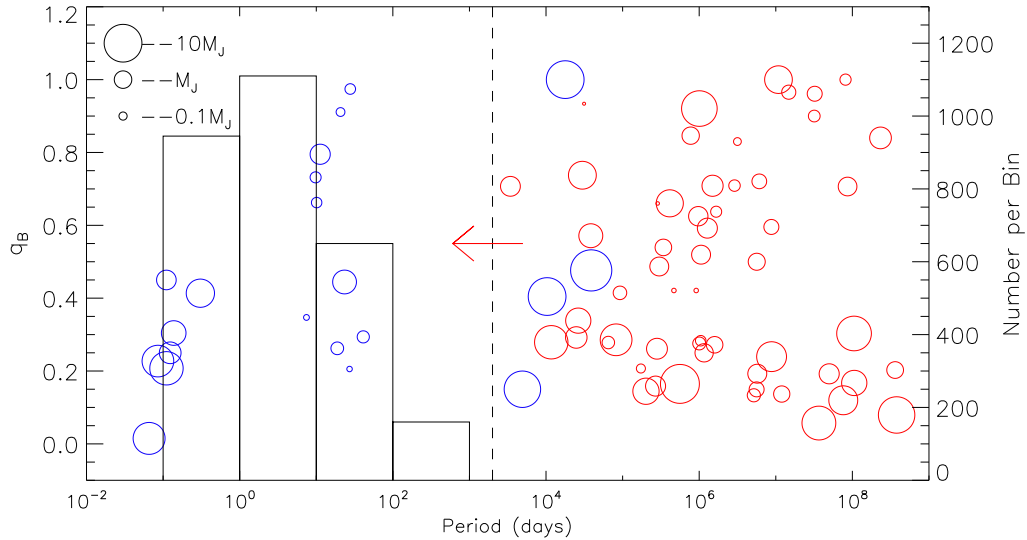}
  \centering
  \caption{The period vs. mass ratio distribution of the binaries which harbour planets. The red circles represent S-type planets, and the blue circles represent P-type planets (circumbinary planets). The size of the circle is proportional to $(m_{p})^{1/3}$. We get the mass of planet on http://www.univie.ac.at/adg/schwarz/multiple.html. For the planets discovered by the RVs, the mass is $m_{p}\sin{i}$. The horizontal axis represents the orbital period of the binary, and the left vertical axis stands for the mass ratio of the binary ($q_{B}=M_{2}/M_{1}$, $M_{1}$ is the mass of the primary). For the multi-planet systems in binaries, we take the innermost planet as a typical body. The histogram exhibits period distribution of the eclipsing binaries discovered by \textit{Kepler} telescope. The right vertical axis is the number of eclipsing binaries per period bin. \textit{Note}: Figure 1 shows only the binaries with the known mass ratio.}\label{fig1}
 \end{figure*}

Recently, a new detection method for S-type planets in eclipsing binaries was proposed by \citet{Oshagh2017}. Using a correlation between the stellar radial velocities (RVs), eclipse timing variations (ETVs), and eclipse duration variations (EDVs), S-type planet can be detected with the existing high-accuracy RV and photometric instruments. \citet{Martin2017} further showed that the orbital precession of S-type planet can raise the transit probability up to as high as tens of per cent. In such case, this will make the detection probability rise up if the binary is known to eclipse. Future space missions, e.g., PLATO (scheduled for launch in 2025), will offer the opportunity to discover S-type planets in eclipsing binaries \citep{Oshagh2017, Martin2017}. Once S-type planets in eclipsing binaries are found via observations, it will significantly improve our understanding of planet formation.

However, according to the traditional theories of planet formation, \textit{in situ} formation of S-type planets in close binaries is very challenging. Firstly, the protoplanetary disk surrounding the host star will be tidally truncated by the companion star \citep{Artymowicz1994,Miranda2015}. As a result, the mass of the disk is remarkably reduced and its lifetime becomes shorter. Secondly, dynamical perturbations from the companion will pump up the eccentricities of planetesimals and enhance their relative collision speed. It is detrimental to the accretion of the planetesimals \citep{Thebault2008, Xie2010}. Thirdly, even if the accretion process can take place in close binaries systems, the accretion timescale will be much longer than that of the single star system. The protoplanetary disk may dissipate before the planet was fully formed. However, the planetary detection in past years infers that planets seem to be ubiquitous. The planetary formation theory does not necessarily deliver ultimate constraints on the actual presence of planets \citep{Oshagh2017}.

Compared with the difficulty in formation of S-type planets in close binaries, it is comparatively easy to yield P-type planets around them. More than 20 planets have been found encircling close binaries (see blue circles in Figure 1). In particular, \textit{Kepler} telescope found a dozen circumbinary planets in main-sequence star binaries. Currently, it is suggested that the selection effect of observations would make the scarcity of CBP in the exoplanet population. \citet{Armstrong2014} suggested that the occurrence rate of circumbinary planets, if they are fairly flat, is comparable to that of planets in single star systems. In addition, many circumbinary gas and debris disks are discovered over the past years \citep{Rodriguez2010,Pietu2011,Kennedy2012}. These observations provide evidence that formation of P-type planets around close binaries would be common in the universe.

Several studies on circumbinary planets suggested that planet-planet scattering can occur in their evolution. \citet{Bromley2015} showed that the free eccentricity of Kepler-34b and Kepler-413b is much greater than their forced eccentricity. The high free eccentricity tends to preclude the migrate-in-gas mode, thereby being consistent with scattering events. \citet{Pierens2008} and \citet{Kley2015} took into account the model of disk-driven migration of multiple circumbinary planets. They showed that planet-planet scattering may naturally result from convergent migration of multiple planets. In the formation scenario, the planets are firstly trapped into a mean motion resonance (MMR) (e.g., 2:1, 3:2, 5:3, 7:4, etc.) configuration. Along with the dissipation of gas disk, the eccentricities of planets can be further stirred up, subsequently the planet-planet scattering will ensue. \citet{Gong2017a} showed that the currently discovered circumbinary planets are more likely to be products of the scattering of multi-planetary systems. \citet{Smullen2016} explored the scattering of multiple circumbinary planets and found that the planet-planet scattering in proximity to binaries leads to more ejections of planets than planet-planet or planet-star collisions in comparison with planet-planet scattering in single star systems.

Moreover, circumbinary planetary systems appear to be dynamically packed. One notable feature of multi-planet systems is that they are dynamically packed \citep{Fang2013,Pu2015}. The single or double planetary systems may be the descendants of more closely packed high-multiple systems \citep{Pu2015}. The eccentricity distribution of planets from radial velocity survey is consistent with expectations for those systems born overpacked, and relaxed to their current configurations through planet-planet scattering \citep{Juric2008}. Such scenario may play a  part in the dynamical evolution of circumbinary planets. \citet{Kratter2014} showed that Kepler-47 system is dynamically packed, although the known population of circumbinary planets is too small to assert whether they are generally a packed population or not. In a word, dynamical scattering may be easily triggered in an overpacked circumbinary planet system.

The previous investigations indicate that the scenario of scattering and tidal capture can throw light on the formation of hot Jupiters in single star systems. This is the so-called high eccentricity migration mechanism \citep{Wu2003,Nagasawa2008,Nagasawa2011,Beauge2012}. In this work, we aim to explore the formation of S-type planets in close binaries by considering the scenario of the scattering of P-type planets and tidal interactions from stars, thereby turning P-type orbit into S-type orbit in close binaries.

\citet{Sutherland2016} treated CBP as test particle and studied the fate of unstable planet. The instability of planet is caused by the n:1 resonance between planet and the inner binary. They showed that a majority of unstable planets will be scattered out of the system. The probability of being captured by inner binary is very low. On observation
CBPs found by \textit{Kepler} are in the resonant cell \citep{Popova2013}, e.g., where Kepler-16b is observed in a resonance cell bounded by the unstable 5:1 and 6:1 mean motion resonances. In general, it is believed that these planets are initially formed distant from the binary and then undergo inward migration in the disk, indicating that CBP can \textit{safely} migrates across these n:1 resonances in the gas disk and ultimately reach their current location. However, as aforementioned, the scattering of the planets will occur when they undergo convergent migration in multi-planets systems. In our work, we model CBPs as massive bodies and consider the scattering scenario in multi-planet systems. Our study shows that scattering can play an effective role in reducing the energy of planets scattered inwards and greatly increase the capture probability of CBP. The capture probability is related to the mass ratio and eccentricity of the binary. Moreover,  we emphasize that  retrograde S-type planets can form through the tidal capture from our simulations. Based on these investigations, our work further provides several suggestions for selecting target binaries to search for S-type planets in the near future.

This paper is organized as follows. In Section 2, we briefly describe the scattering induced tidal capture scenario in close binaries, and then we present the model of this work. In Section 3, we present the initial conditions and numerical results.  Finally, Section 4 summarizes the major results and compares our results with those of high-eccentricity migration mechanism in single star systems.

\section{Tidal capture scenario and simulation model}

We first describe the scattering induced tidal capture scenario with the aid of the planar circular restrictive three-body problem (CRTBP). There is an unstable boundary around binary, beyond which circumbinary planets can form and exist. \citet{Holman1999} derived this boundary by numerical simulations,
\begin{equation}
\begin{array}{l}
{a_{c,out}} = \left[ {1.6 + 4.12\mu  + 5.1{e_B} - 4.27\mu {e_B} - } \right.\\
\left. {\quad \quad \,\;\;2.22e_B^2 - 5.09{\mu ^2} + 4.61{\mu ^2}e_B^2} \right]{a_B}
\end{array}
\end{equation}
where $\mu  = {{{M_2}} \mathord{\left/
 {\vphantom {{{M_2}} {\left( {{M_1} + {M_2}} \right)}}} \right.
 \kern-\nulldelimiterspace} {\left( {{M_1} + {M_2}} \right)}}$  is the mass ratio of the binary, ${M_2}$ is the mass of the less massive star. $a_B$ and $e_B$ is the SMA and eccentricity of the binary, respectively.

If CBP is regarded as a test particle, from the point of view of CRTBP, there exists a Jacobi constant satisfying the condition $C_{J} > C_{J}(L_{2})$ for orbits to remain stable\footnote{For $C_{J} > C_{J}(L_{2})$ the zero velocity surfaces delimit three regions where the motion of
the planet is possible. One is the exterior of the binary, and the other two are closed around the primary and the secondary, respectively.}. Herein $C_{J}(L_{2})$ denotes a critical value of $C_{J}$, which is relevant to Lagrange point $L_{2}$ \citep{Murray1999}. It is impossible for the planet to penetrate the forbidden area around the binary and approach either of the stars . However, in a multiple CBP system, $C_{J}$ of a planet is not perfectly conserved when one considers the gravitational perturbations of the other planets. Especially the change of $C_{J}$ is significant during close encounters with other planets.  Planet-planet scattering can bring about a sudden decrease in $C_{J}$, to make $C_{J} < C_{J}(L_{2})$. When this condition is satisfied, the planet can approach either of the binary. If the planet is scattered far enough away from the other planets, then its $C_{J}$ might maintain a value on appreciable time-scales.  Again, the motion of the scattered planet can be dictated by the CRTBP as long as the perturbations from the other planets are negligible. If we do not consider other dissipations such as tides, the scattering can lead to a \textit{temporary capture}. In most cases planets will be scattered out of the system or collide with two stars \citep{Gong2017b}. However, tidal interaction from the star becomes more important when the periastron of the planet is very close to either of binary. Therefore, tidal effects between the planet and the star will further diminish the orbital energy of the planet, thereby producing $C_{J} > C_{J}(L_{2})$. Under such circumstance, a \textit{permanent capture} of the planet can form. Figure 2 shows an example of this scenario using an artificial two-planet system. In the following, we will briefly introduce our model.

\begin{figure*}
\includegraphics[scale=0.70]{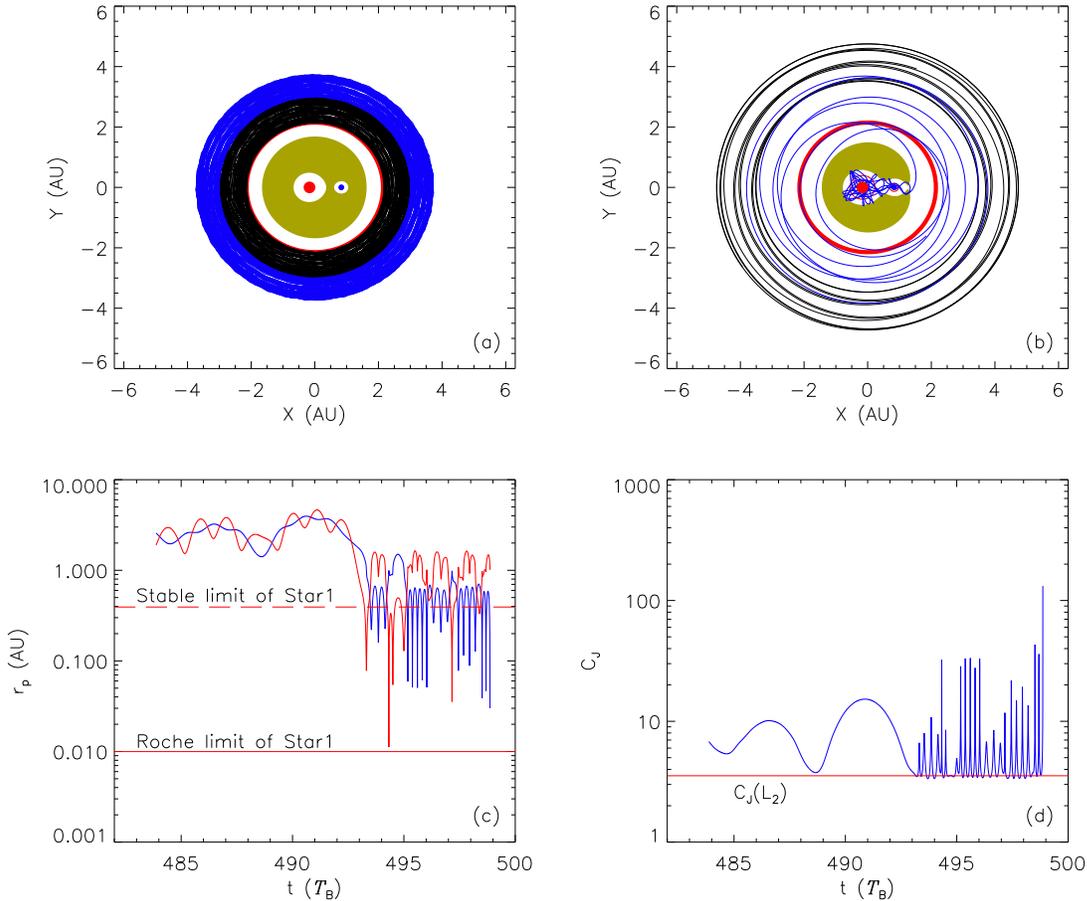}
  \centering
  \caption{An example of scattering induced tidal capture of circumbinary planet. The parameters of the binary are: $M_{1}=M_{\odot}$, $M_{2}=0.2M_{\odot}$, $a_{B}=1$ au, $e_{B}=0$. (a) Two circumbinary planets are modeled as test particles. Their orbits are long-term stable. For clarity, only the final stage of evolution is shown in all panels. The red dot and the blue dot represent the primary and the companion star, respectively. The thick red line is the unstable boundary outside the binary. The brown regime is the zero velocity surface of $C_{J}=4$ shown for reference. (b) The two planets have identical initial conditions as in Figure (a), but the scattering between two planets has been considered. The initial outer planet is tidally captured as an S-type planet by the primary star. The red circles around two stars are each of their unstable boundaries. (c) Red line is time varying distance between the captured planet and the secondary. The blue line shows the evolution of the distance of the captured planet from the primary. The red dotted line is the stable boundary of the primary, and the red solid line is its Roche's limit. (d) The Figure shows how the Jacobi constant of the captured planet changes with time. The red solid line denotes the value of ${C_J}\left( {{L_2}} \right)=3.53$.
  }\label{fig2}
\end{figure*}

\subsection{Tides}

Currently, two kinds of tide models are constructed to understand the planetary evolution for two limit cases: equilibrium tide \citep{Mignard1980,Ferraz2008} and dynamical tide \citep{Ivanov2004,Ivanov2007}. The equilibrium tide is suitable for low eccentricity orbit of the planet, whereas the dynamical tide can be applied to the cases where the eccentricities of the planetary orbits are close to 1. When a scattered planet approaches either star of the binary, it travels on a parabolic or high-eccentricity orbit with respect to the star. Therefore, the dynamical tide model is suitable for capture process. Our numerical results showed that the planets still remain on high-eccentricity orbits after capture.  They will undergo a long timescale of tidal evolution as does in single star system. As this work mainly focuses on the capture probability of planets, we do not model the evolution of S-type planets after capture.

However, the dynamical tides are much more complex than equilibrium tides. Currently, the mechanisms that how the dynamical tides affect the orbital and rotational evolution of the participating bodies are not very clear \citep{Beauge2012,Wu2018}. Moreover, in the analytical form \citep{Beauge2012}, the above two tidal models do not take into account the mass loss of planet. If they pass approximate to or within Roche limit of a star at periastron, the planets will suffer from mass loss. In order to reflect the roles of the dynamical tide and the tidal disruption of planets, we adopt the tide model given in \citet{Faber2005} (see Figure 3), where the tidal interactions and disruption of giant planets with highly eccentric orbits were explored in their work.

Further studies on the role of a dense core of giant planets in the tidal disruption process was discussed in \citet{Liu2013}. For a low core mass, their outcome was similar to that of \citet{Faber2005}. Herein we utilize impulse approximation to model tide dissipation \citep{Nagasawa2008, Nagasawa2011}. In this model, the relative velocity of the planet with respect to the star is changed discontinuously at the pericenter passage described as
\begin{equation}
\boldsymbol{v}' = \sqrt {2\frac{{\Delta {E_{orb}}\left( {{r_p}} \right)}}{{{m_p}}} + {v^2}} \frac{\boldsymbol{v}}{v},
\end{equation}
where $\Delta {E_{orb}}\left( {{r_p}} \right)$ changes according to the law showed in Figure 3, and ${r_p}$ denotes the distance between the planet and the star at its closest approach.

In all cases we calculated, the mass of the planet is ${m_p} = {m_J}$. In addition, we allow the mass of planets to vary according to the law in Figure 3 when they suffer from mass loss.
\begin{equation}
{m'_p} = {m_p}\left[ {1 - \frac{{\Delta {m_J}\left( {{r_p}} \right)}}{{{m_J}}}} \right]
\end{equation}
Certainly, the tide model will be inapplicable when the mass loss of planets is considerable. However, our numerical simulations show that most planets do not suffer any mass loss at the time of capture. In our model the mass of the primary is set to be 1 ${M_ \odot }$. The mass of the companion star is ${M_2} = {q_B}{M_1}$, where ${q_B}$ is the mass ratio of the binary. In our simulations, we take ${q_B} = 0.1$, 0.3, 0.5, 1.0, respectively. As ${M_2}$ changes, the tidal interaction between the companion and the planet  changes accordingly. We make a linear approximation for the orbital energy change based on the Equation (7) in \citet{Nagasawa2008}, that is, $\Delta {E'_{orb}} \approx \sqrt {{M_2}} \Delta {E_{orb}}$, where $\Delta {E_{orb}}$ is for the primary.

\subsection{How to define a capture}

According to \citet{Holman1999}, there is a stable boundary encircling each star of a binary, within which the planet orbit is stable.
\begin{equation}
\begin{array}{l}
{a_{c,in}} = \left[ {0.458 - 0.39\mu  - 0.655{e_B} + 0.525\mu {e_B} + } \right.\\
\left. {\quad \quad \;1.09e_B^2 - 0.272\mu e_B^2} \right]{a_B},
\end{array}
\end{equation}
where $\mu  = {{{M_2}} \mathord{\left/
 {\vphantom {{{M_2}} {\left( {{M_1} + {M_2}} \right)}}} \right.
 \kern-\nulldelimiterspace} {\left( {{M_1} + {M_2}} \right)}}$ for the primary, and $\mu  = {{{M_1}} \mathord{\left/
 {\vphantom {{{M_1}} {\left( {{M_1} + {M_2}} \right)}}} \right.
 \kern-\nulldelimiterspace} {\left( {{M_1} + {M_2}} \right)}}$ for the secondary. We use this boundary to determine the capture. If a planet is on the bounding orbit (${E_{orb}} < 0$) with respect to a star and its aphelion satisfies $Q = {a_p}\left( {1 + {e_p}} \right) < {a_{c,in}}$, we consider that the planet is captured by the star. When the above-mentioned conditions are satisfied, we mark the planet as `capture' and remove it from the system.

 At the same time, we further examine whether the periastron of the captured planet is larger than the Roche limit of the star, with $q = {a_p}\left( {1 - {e_p}} \right) > {r_R}$. Our numerical results show that most of the captured planets can satisfy $q = {a_p}\left( {1 - {e_p}} \right) > {r_R}$. A small fraction of them have $q = {a_p}\left( {1 - {e_p}} \right) < {r_R}$. This portion of planets will undergo mass loss during the subsequent post-capture evolution. In this case, their ultimate fate is unknown - either they may still remain in the bounded orbit, or they would escape from their host star due to significant mass loss \citep{Faber2005,Guillochon2011,Liu2013}. However, this is beyond the scope of this work and herein we will not discuss their long-term post-capture evolution.

\subsection{The codes}
In this work, we perform extensive numerical integrations using MERCURY package \citep{Chambers1999}.
To simulate the dynamical evolution of circumbinary planets, \citet{Smullen2016} modified MERCURY package
and made it suitable for close binaries. The codes (MERCURY\_RAS) had been well examined in their work and is publicly available (https://github.com/rsmullen/mercury6\_binary). Herein, we adopt this update MERCURY package and incorporate above-mentioned tide model into the package by modifying the codes.

MERCURY\_RAS treats close encounters between any pair of bodies in the same way, in contrast to the standard MERCURY which deals with encounters with the central star separately. For any star-planet pairs, MERCURY\_RAS searches for close encounters between them. The close encounter radius of a star is set to be $n \cdot R_{S}$ in this work, where $R_{S}$ is the stellar radii and $n$ is a real number. We take $n \cdot R_{S}=4.0~r_{t}$, where $4.0~r_{t}$ is the range of tidal effect as shown in Figure 3. Subsequently, we added the tidal interactions in the codes by updating the subroutine of MCE\_STAT.FOR of MERCURY\_RAS. When each star-planet pair has a close encounter, the subroutine uses an interpolation method to estimate the minimum distance between two bodies (namely pericentre passages). In our revision, we change the velocity of planet at pericentre passages according to Equation (2). Therefore, the modified codes are employed for our simulations. The code test is described in Appendix A.

\section{Numerical simulation and analysis}

\subsection{The initial setup}
Planet-planet scattering is widely known as one of the major mechanisms that shed light on the formation of hot Jupiters in single star systems. These previous studies generally consider the scattering of three giant planets \citep{Nagasawa2008,Chatterjee2008,Nagasawa2011,Beauge2012}. In order to compare the capture efficiency in single star systems with that in close binaries, we study the scattering scenario of three circumbinary planets in the systems. An actual exoplanet system is the Kepler-47(AB) bcd system \citep{Orosz2012, Hinse2015}. In this work, we assume the masses of three planets are all identical to Jupiter mass. The investigation of disk-driven migration of CBP showed that the planet will migrate inwards and eventually be stalled near the unstable boundary of the binary. From the observations, we learn that most CBPs discovered by \textit{Kepler} do cluster simply outside of this boundary. Therefore, we adopt SMA of the innermost planet to be 1.1 $a_{c,out}$ in the simulations.

The initial spacing between planets affects the unstable timescale of the investigated system. Herein we set the initial spacing of three planets according to the hydrodynamical simulations of multi-CBP system \citep{Pierens2008,Kley2015}. The previous studies show that CBPs will be captured into the orbital resonance configuration in circumbinary disk, such as 2:1, 3:2, 5:3 MMRs and so on. In the multi-planet systems reported by \textit{Kepler}, 2:1 and 3:2 (or near) resonances are observed to be most common amongst plenty of single star systems \citep{Wang12,Petrovich2013,Fabrycky2014,Wang2014,Marti16,Sun2017}. Interestingly, recent investigation unveils that planet pairs have a higher likelihood to be trapped into 3:2 MMR than 2:1 MMR if the scenarios of mass accretion of planets and potential outward migration are taken into account \citep{Wang2017}. Therefore, in this work we take the initial spacing of the adjacent planets greater than 3:2 MMR but less than 2:1 MMR. The spacings of other resonances such as 5:3 or 7:4 are between 3:2 and 2:1 resonances. To alleviate the computational burden, we do not explore the cases where the initial spacing of planets is very large. The initial SMAs of three planets are described as follows.
\begin{equation}
{a_{i + 1}} = {a_i} + K \cdot {R_{Hill,m}}\quad \;\left( {i = 1,\,2} \right),
\end{equation}
\begin{equation}
{R_{Hill,m}} = {\left( {\frac{{{m_i} + {m_{i + 1}}}}{{3{M_ * }}}} \right)^{{1 \mathord{\left/
 {\vphantom {1 3}} \right.
 \kern-\nulldelimiterspace} 3}}}\left( {\frac{{{a_i} + {a_{i + 1}}}}{2}} \right),
\end{equation}
where ${R_{Hill,m}}$ is the mutual Hill radius of the adjacent planets. We adjust the value of $K$ to make the initial spacing of planets is greater than 3:2 MMR but less than 2:1 MMR. The initial eccentricity of the planet is $\le {10^{ - 3}}$. Planets and the binary are in a coplanar configuration. We also assume that the planets are initially on the prograde orbit (relative to both the rotation of each star and the revolution of the binary).  The three phase angles of the planets' orbits are chosen randomly and uniformly ranging from 0 to 360 degrees. We simulate the binaries with three different SMA of $a_{B}$=1 au, 0.5 au, and 3 au \footnote{\citet{Trilling2007} found circumbinary discs are common around binaries with separations less than 3 au.}, respectively. For each $a_{B}$, we consider variational mass ratio $q_{B}$ and eccentricity $e_{B}$ of the binary. For $e_B=0$, we take $q_B$=0.1, 0.3, 0.5, 1.0. And for $q_B$=0.1, we explore $e_B$=0.0, 0.1, 0.3, 0.5. For each binary configuration, we perform 1000 runs to investigate the capture scenario and scattering evolution. Our analysis presented in this work is based on over 21,000 runs for various cases of configurations of circumbinary planets.

\begin{figure*}
\includegraphics[scale=0.70]{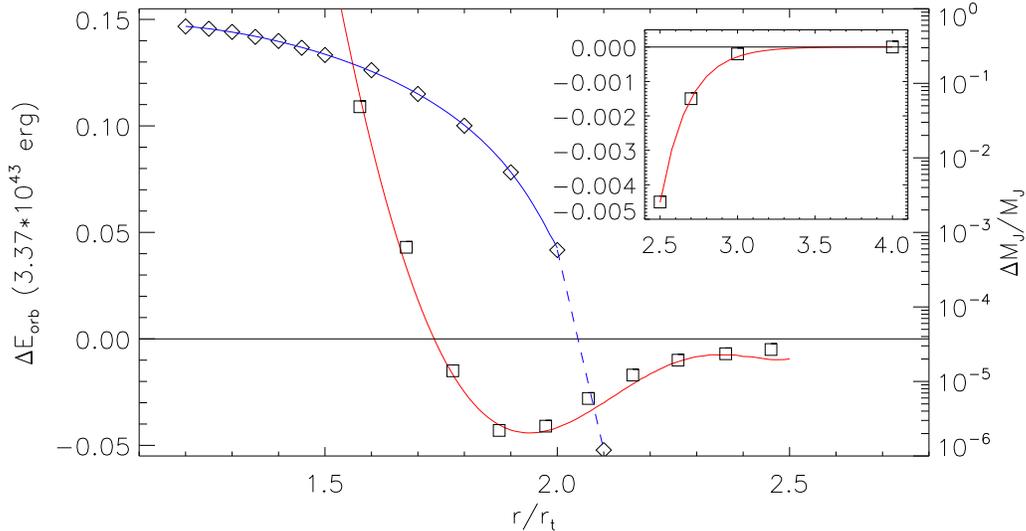}
  \centering
  \caption{The tide model used in this work. We piecewise fit the results of \citet{Faber2005} (Table 1 in their paper) . The horizontal axis is the distance from the star at the closest approach in the unit of tidal radius $(M_{\ast}/m_{p})^{1/3}R_{p}$. The red line shows the orbital energy change (left vertical axis) of the planet as it passes through the periastron. For $r<2.5$, we use polynomial fitting, whereas Gaussian fitting is adopted for $2.5 < r < 4.0$. The blue line shows the change in mass (right vertical axis) of the planet as it passes through the periastron. The solid blue line is polynomial fitting, and the dotted blue line is a linear fit.}
 \label{fig3}
\end{figure*}

\subsection{Results and analysis}

We set $a_{B}$=1 au as the fiducial binary configuration and analyze the simulation results in detail. The results are summarized in Table 1. In the table, $Total$ represents the entire capture percentage out of 1000 runs. $S_{1cap}$ stands for the percentage of planets captured by the primary, whereas $S_{2cap}$ is the percentage of planets captured by the companion star. $Retro.$ is the percentage of capture that have produced retrograde planets, and $T_{dis}$ is the fraction of cases in which the planet was captured but with mass loss. For $e_{B}=0$, our results show that the probability of capture becomes smaller as the mass ratio increases. For $q_{B}=0.1$, the total fraction of capture is $12.8\%$, which can be comparable to the occurrence
rate of hot Jupiters through planet-planet scattering in single star systems \citep{Beauge2012}. In the work of \citet{Beauge2012}, the formation rate of hot Jupiters is $\sim10\%$. Hence, we draw a conclusion that this mechanism of tidal trapping will play a vital part in the scattered circumbinary planet of close binaries with a suitable $q_{B}$ as in single star systems.

However, the entire capture fraction becomes smaller as the mass ratio increases. When the mass ratio is ${q_B} = 1$, the capture fraction is simply $1.4\%$, which is much smaller than the case of ${q_B} = 0.1$ by one order of magnitude. In our simulations, we found that two parameters can play an essential role in the capture efficiency: $a_{1,0}$ and $C_{J}(L_{2})$, which are the function of ${q_B}$  (see Figure 4). As can be noted in Figure 4, the blue line represents ${a_{1,0}}$  as a function of ${q_B}$, while the red profile shows $C_{J}(L_{2})$ changes with ${q_B}$. In the case of ${q_B} = 1$, i.e., ${M_1} = {M_2} = {M_ \odot }$, the perturbation of the binary becomes strongest. The unstable boundary, as well as the initial position of the planets, is most distant from the binary. However, $C_{J}(L_{2})$ has a minimum value at ${q_B} = 1$. The zero velocity surface of CRTBP shows that planet has a chance to be captured by the binary only if the condition $C_{J}<C_{J}(L_{2})$ is satisfied. When the regime of planet-planet scattering is far away from the binary, the scattering between CBPs cannot give rise to enough energy change to efficiently drop the $C_{J}$ of the scattered (inward) planet. As a result, the capture probability drops significantly for equal-mass binary.

For the case of $e_{B}=0$, both the primary and the secondary can seize the planet. The capture fraction of the companion ($9.3\%$) is larger than that of the primary ($3.5\%$) for $q_{B}=0.1$. However, when $q_{B}=0.3$, the capture probability of the primary ($4.1\%$) is larger than that of the companion ($1.5\%$). We speculate that this tendency may be induced by the variations of $a_{1,0}$ and $C_{J}(L_{2})$ as shown in Figure 4. From the panels of Figure 2(b) and 2(d), we note that the captured planets mainly enter the initial limited area of the binary through the vicinity of $L_{2}$ point. That is, it first passes near the companion and then moves to the vicinity of the primary if the orbital energy is large enough. However, $C_{J}(L_{2})$ is lower when $q_{B}=0.1$. Planet-planet scattering can provide enough energy to drive the scattered planet pass through $L_{2}$ to the outskirt of the companion star, but cannot make it approach the primary. In this sense, the capture probability of the companion is greater than the primary. As $q_{B}$ increases, $C_{J}(L_{2})$ increases as well. The energy change caused by the scattering makes the planet meet the capture condition ($C_{J}<C_{J}(L_{2})$) readily. Planet has enough energy to approach the primary. On the other hand, the stable area encircling the primary is broader than that of the companion. Therefore, the capture probability of the primary star rises up. The capture likelihood of a planet (by $M_{1}$ or $M_{2}$) depends on its bouncing timescale between two stars \citep{Moeckel2012} and their tidal dissipation efficiency.

In all binary configurations we explored (see Table 1-3), there exist the resultant planets with retrograde orbits. As for $e_{B} = 0$, the formation percentage decreases as $q_{B}$ goes up. Similarly, when $q_{B}$ keeps constant, the formation probability of the retrograde planets falls when $e_{B}$ increases. An paradigm is illustrated in Figure 5. The understanding of origin of a retrograde planet can offer a key clue to the formation of circumstellar planets. For the coplanar configurations we account for, if S-type planets form locally in a circumstellar disk accompanying with the primary or secondary, they should move on prograde orbits. However, the retrograde orbit can only be shaped through a dynamical capture in the system. As a consequence, the retrograde orbit can be used as a fingerprint to distinguish between \textit{in situ} formation and post-capture scenario. The observations provide evidence that several S-type planets are thought on the retrograde orbits, such as $\nu$ Octantis \citep{Eberle2010,Ramm2016}, HD 59686 Ab \citep{Ortiz2016,Trifonov2018}. HD 59686 Ab is an S-type giant planet discovered in a close ($a_{B}$ = 13.6 au) and eccentric ($e_{B}$ = 0.73) binary. It is very likely to be on a coplanar and retrograde orbit based on stability analysis \citep{Trifonov2018}. The two stars are separated by only 3.6 au at periastron. Hence, it is very difficult to produce HD 59686 Ab through the scenarios of in situ formation by core accretion or disk instability \citep{Ortiz2016}. Especially, forming a retrograde planet requires exotic scenarios \citep{Trifonov2018}. The capture may act as one of likely mechanisms to elucidate the formation of these planets.

From our simulations, we found most of captured planets did not undergo tidal disruption in the dynamical evolution. As a matter of fact, only a very small proportion of captured planets suffer from mass loss. The mass loss rate ${{\Delta {m_p}} \mathord{\left/
 {\vphantom {{\Delta {m_p}} {{m_p}}}} \right.
 \kern-\nulldelimiterspace} {{m_p}}}$ varies accordingly for various $e_{B}$ and $q_{B}$.  Taking $e_{B} = 0$ and $q_{B} = 0.1$ as an example, 14 out of 1000 runs experience mass loss. The maximum mass loss is 0.604, whereas the minimum value is 3.67$\times 10^{-4}$. And the median value of mass loss is 4.71$\times 10^{-2}$. As aforementioned, the tide model we used will be no longer appliable with a large mass loss of planet. However, it has little effect on the overall capture probability of planets due to their low occurrence rate.

\begin{figure}
\includegraphics[scale=0.70]{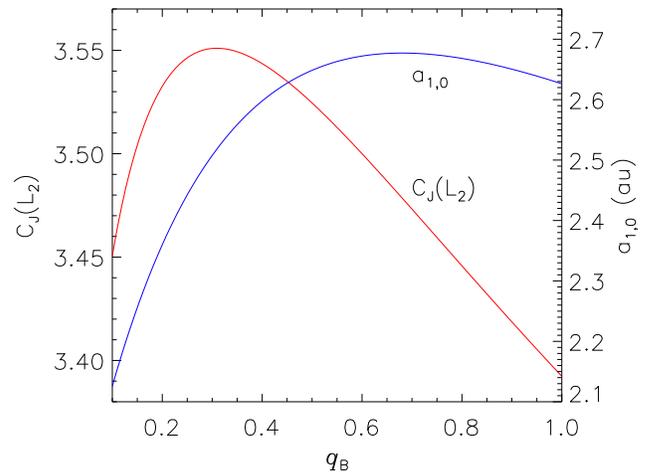}
  \centering
  \caption{The red line exhibits the Jacobian constant ${C_J}\left( {{L_2}} \right)$ as a function of the mass ratio of the binary (the left vertical axis). The blue line is the initial semi-major axis of the innermost planet (the right vertical axis). The mass of the primary is assumed to be 1 $M_{\odot}$. The eccentricity of the binary is 0.
  }\label{fig4}
\end{figure}

For $q_{B}=0.1$, we investigate how the capture probability varies with $e_{B}$. As we can see from Table 1, the total capture probability declines as $e_{B}$ increases. A notable feature is that the capture probability of the secondary is always greater than that of the primary for all cases of $e_{B}$ we explored. It is very clear that there is no Jacobi-integral in the elliptic restricted three-body problems. In a non-uniformly rotating and pulsating coordinate system, the shape and dimension of zero velocity surfaces vary with time. However, the planet can still move into the binaries from the instantaneous `$L_{2}$' point. As we can see in Figure 6, the initial position of planets (unstable boundary of binary) increases monotonically along with $e_{B}$. The scattering takes place farther away from the binary in an eccentric binary case, so the scattered planet does not have enough energy to approach the primary. Planets will be more easily seized by the secondary, which is closer to the instantaneous `$L_{2}$' point.

\begin{figure}
\includegraphics[scale=0.70]{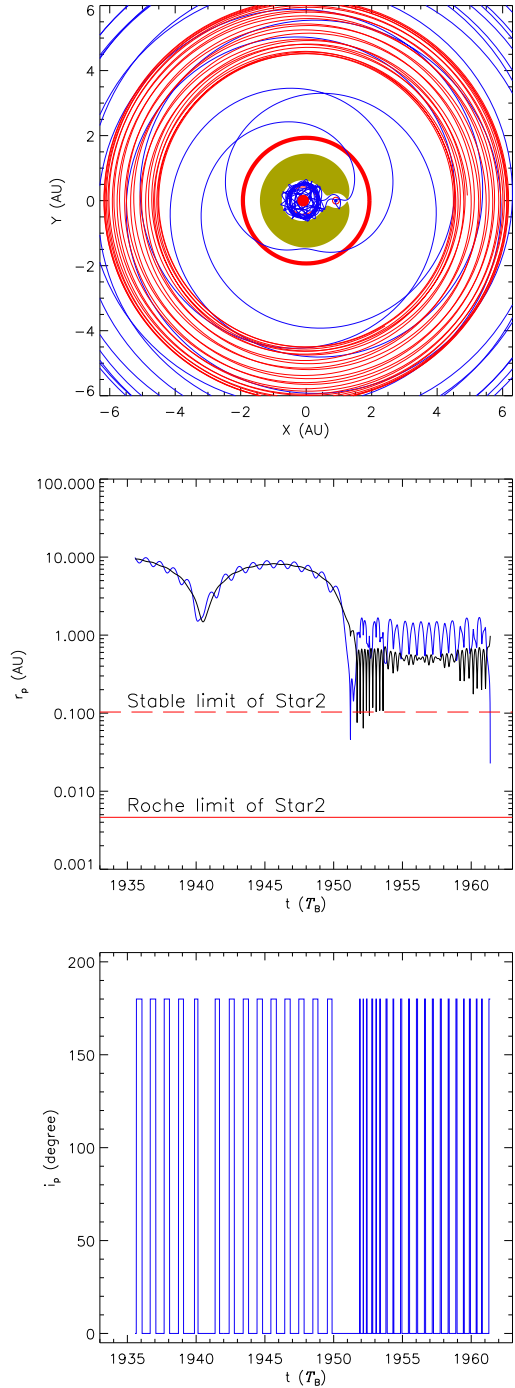}
  \centering
  \caption{An example of retrograde planet formation. There are initially three planets in the system. For clarity, only the evolution of the late stage is shown. One planet is scattered out of the system in much earlier time, and is not shown in the panel. Top panel: the planet in blue is captured by the companion star and moves on the retrograde orbit eventually. Middle panel: the time evolution of the distance of the captured planets from the primary (black) and the companion star (blue). Bottom panel: the dynamical behavior of the inclination of the captured planet relative to the companion star (two-body approximation). After capture, the planet is moving on a retrograde orbit ($180^{\circ }$).
  }\label{fig5}
\end{figure}

\begin{table}
 \begin{minipage}{80mm}
  \caption{Percentages of different outcomes of tidal capture for $a_{B}=1$ au. Total: Fraction of total captures by the inner binary. S$_{1cap.}$: Fraction of total captures by the primary of the binary. S$_{2cap.}$: Capture by the secondary. Retro.: Fraction of cases in which a retrograde planet formed. T$_{dis.}$: Planets suffered from tidal disruption.}
   \begin{tabular}{lllllll}
  \hline
    & q$_B$   &  Total     & S$_{1cap.}$   &   S$_{2cap.}$  &  Retro.  & T$_{dis.}$\\
  \hline
  e$_{B}=0.0$ & 0.1    &   12.8  &   3.5 &   9.3  &    3.2  &  1.4   \\
  & 0.3    &   6.0  &   3.1  &  2.9  &  1.1   &  0.7   \\
  & 0.5    &  4.1  & 2.7  & 1.4   &  0.3  &  0.7   \\
  & 1.0    &  1.4  & 0.7  & 0.7   &  0.3  &  0.0   \\
  \hline
  &e$_B$   &  Total     & S$_{1cap}$   &   S$_{2cap}$  &  Retro.  & T$_{dis.}$\\
  \hline
  q$_{B}=0.1$   & 0.0    &   12.8  &   3.5 &   9.3  &    3.2  &  1.4   \\
  & 0.1    &   10.3  &   3.6  &  6.7  &  2.7   &  0.8   \\
  & 0.3    &   6.5   & 2.4  & 4.1   &  2.2  &  0.6   \\
  & 0.5    &   3.5   & 1.2  & 2.3   &  1.1  &  0.6   \\
\hline
\end{tabular}
\end{minipage}
\end{table}

\begin{table}
\begin{minipage}{80mm}
  \caption{Percentages of various outcomes of tidal capture for $a_{B}=0.5$ au}
   \begin{tabular}{lllllll}
  \hline
    & q$_B$   &  Total     & S$_{1cap.}$   &   S$_{2cap.}$  &  Retro.  & T$_{dis.}$\\
  \hline
  e$_{B}=0.0$ & 0.1    &   11.4  &   4.8 &   6.6  &    2.3  &  1.4   \\
              & 0.3    &   5.6  &   4.1  &  1.5  &  0.7   &  0.7   \\
              & 0.5    &  2.6  & 1.5  & 1.1   &  0.1  &  0.6   \\
              & 1.0    &  0.7  & 0.3  & 0.4   &  0.1  &  0.1   \\
  \hline
  &e$_B$   &  Total     & S$_{1cap}$   &   S$_{2cap}$  &  Retro.  & T$_{dis.}$\\
  \hline
  q$_{B}=0.1$   & 0.0    &   11.4  &   4.8 &   6.6  &    2.3  &  1.4   \\
                & 0.1    &   9.5  &   3.8  &  5.7  &  2.1   &  1.7   \\
                & 0.3    &   4.0   & 1.7  & 2.3   &  1.1  &  0.6   \\
                & 0.5    &   2.0   & 0.8  & 1.2   &  0.5  &  0.3   \\
\hline
\end{tabular}
\end{minipage}
\end{table}

\begin{table}
 \begin{minipage}{80mm}
  \caption{Percentages of different outcomes of tidal capture for $a_{B}=3$ au}
   \begin{tabular}{lllllll}
  \hline
    & q$_B$   &  Total     & S$_{1cap.}$   &   S$_{2cap.}$  &  Retro.  & T$_{dis.}$ \\
  \hline
  e$_{B}=0.0$ & 0.1    &   10.7  &   2.4 &   8.3  &    3.0  &  1.1   \\
              & 0.3    &   4.4  &   2.0  &  2.4  &  0.9   &  0.1   \\
              & 0.5    &  3.9  & 1.7  & 2.2   &  0.7  &  0.3   \\
              & 1.0    &  1.2  & 0.6  & 0.6   &  0.4  &  0.0   \\
  \hline
  &e$_B$   &  Total     & S$_{1cap}$   &   S$_{2cap}$  &  Retro.  & T$_{dis.}$\\
  \hline
  q$_{B}=0.1$   & 0.0    &   10.7  &   2.4 &   8.3  &    3.0  &  1.1   \\
                & 0.1    &   6.0  &   1.7  &  4.3  &  1.9   &  0.9   \\
                & 0.3    &   4.8   & 1.6  & 3.2   &  1.6  &  0.3   \\
                & 0.5    &   3.0   & 1.1  & 1.9   &  1.4  &  0.6   \\
\hline
\end{tabular}
\end{minipage}
\end{table}

We further explored the case of $a_{B} = 0.5$ au and $a_{B} = 3$ au (see Tables 2 and 3), and we draw similar conclusions. The maximum capture probability is approximately  $10\%$ for low binary mass ratio. In comparison with the case of $a_{B}=1$, the entire capture probability decreases slightly for the cases of $a_{B}=0.5$ and 3 au, respectively. For the very close binary, the stable region of the component star shrinks, so the capture probability drops. For a larger $a_{B}$, the required timescale for a system to fully evolve will take longer time. As we stop numerical integrations at $10^6$ years for all cases of $a_{B}$, the capture probability appears to be reduced for the case of $a_{B}=3$ au. Principally, for a larger $a_{B}$, the stable regime surrounding the component star will be broader so that it is much easier for the planet to be caught by the star, thereby enhancing its capture probability.

\begin{figure}
\includegraphics[scale=0.70]{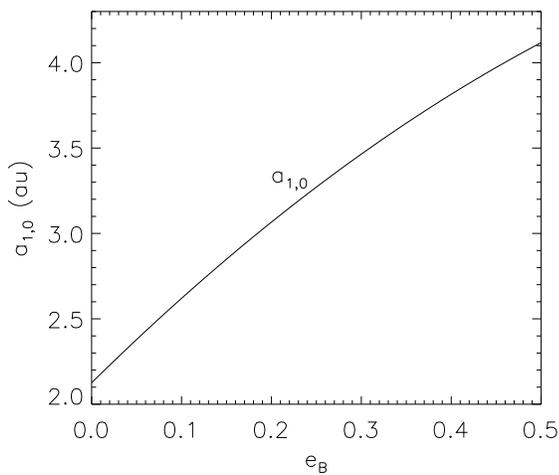}
  \centering
  \caption{The initial semi-major axis of the innermost planet as a function of the eccentricity of the binary. The mass of the primary and the secondary are 1 $M_{\odot}$ and 0.1 $M_{\odot}$, respectively. The semi-major axis of the binary is 1 au.
  }\label{fig6}
\end{figure}

\section{Summary and discussion}

Recent observations have reported a number of S-type planets in binary systems.  However, S-type planets have not yet been found in close binaries with $a_{B}<5$ au, although \textit{Kepler} mission has revealed over 3000 eclipsing binaries with $P < 1000$ days. In the forthcoming space missions such as PLATO \citep{Oshagh2017, Martin2017}, along with the development of high-accuracy RV and high-resolution photometric measurements, S-type planets in close binaries are hopefully to be discovered. From a viewpoint of planetary formation, the \textit{in situ} formation of S-type planets in close binaries seems to be difficult. In this work, we extensively carry out numerical simulations to explore the scenarios of planet-planet scattering amongst circumbinary planets (three P-type giant planets) and subsequent tidal capture process in distinct variety of binary configurations, to examine whether the mechanism can produce such kind of planets. The major conclusions are summarized as follows:

1. The scattering induced tidal capture mechanism can turn a P-type orbit into an S-type orbit. S-type planets in close binaries with SMA of $0.5-3$ au can be yielded through this capture scenario.

2. The formation probability varies with the mass ratio and eccentricity of the binary. The smaller the binary mass ratio, the greater the capture probability. The capture probability is larger in binaries with a small eccentricity than that with a large one.

3. The maximum capture probability is more than $10\%$. The capture probability is approximately identical to the probability of forming hot Jupiters in single-star systems through planet-planet scattering plus tidal capture scenario \citep{Beauge2012}

4. Retrograde S-type planets can be generated through such tidal capture mechanism. The retrograde orbits of  circumbinary planets can be employed as an important indicator that distinguishes in situ formation and post-capture formation scenario.

In single star system, planet-planet scattering, as one of the major mechanisms, can play a crucial role in explaining formation of hot Jupiters. \citet{Chatterjee2008} considered scattering scenarios of three giant planets in single star system without consideration of tidal effect, and revealed that $\sim 10\%$ of the systems harbor planets with periapse distances $<0.15$ au, whereas a few ($\sim 2\%$) harbor planets with periapse distances $<0.03$ au. Therefore, the above probability can be considered as the probability of forming 'potential' hot Jupiters. In the work of \citet{Nagasawa2008} and \citet{Nagasawa2011}, the formation probability of hot Jupiters can amount up to $20-30\%$, which is much higher than those of previous studies \citep{Weidenschilling1996,Marzari2002,Chatterjee2008}. Such high formation probability is mainly triggered by two factors. Firstly, they used dynamical tide throughout the whole orbital circularization. As aforementioned, once the eccentricity of planet is lower than a certain value, the strong dynamical tide is no longer applicable. As a result, this will definitely enhance the formation probability of hot Jupiters. Secondly, the authors pointed out that Kozai mechanism in outer planets during three-planet orbital crossing is also responsible for the formation of close-in planets. Furthermore, \citet{Beauge2012} proposed an empirical formula of tide effects to evaluate both the equilibrium tide and dynamical tide.  The formation probability of hot Jupiters in their work is approximately $10\%$ \citep{Beauge2012}. From the observations, the occurrence rate of hot Jupiter (period $<10$ days; mass $> 0.1$ $M_{J}$) in RV survey is $\sim 1\%$. In the transit surveys the occurrence rate is $\sim 0.1-0.4\%$, the exact numbers slightly differ for different surveys ($\sim 0.4\%$ in \textit{Kepler} mission) \citep{Wright2012,Wang2015}. In our model, the maximum formation probability of S-type is $12.8\%$. Therefore, we conclude that the probability of tidal capture of circumbinary planet is considerable, and this mechanism is valid in close binary systems as in single star systems.

In present work, we simply limit ourselves on the exploration of Jupiter mass planets. Amongst currently discovered $\sim 20$ circumbinary planets, about half of the population have been identified to bear Jupiter masses. In the Kepler CBPs, Kepler-1647b, with a mass of 1.5 $M_{J}$, is a Jupiter-like planet, while Kepler-453b has a mass of 0.03 $M_{J}$, which can be catagorized as a Neptune-like planet. The remaining bodies have planetary masses ranging from 0.13 to 0.53 $M_{J}$, in resemblance to Saturn. The Jupiter mass planets simply occupy the occurrence rate of $\sim 10\%$ of single stars \citep{Cumming2008}. Similarly, for circumbinary planets, Jupiter mass planets appear to be less common \citep{Armstrong2014} as compared to single star systems. Therefore, our study, on the basis of tidal interactions between Jupiter-like planets and their host stars \citep{Faber2005,Liu2013}, cannot be applied to estimate the occurrence rate of S-type planet in close binaries. In this sense, the difficulty in more exploration is that the nowadays dynamical tide model is not suitable for Saturn-like or Neptune-like planets any more. In particular, the inner structure of Neptune-like planets differs a lot from that of gas giants. This may eventually give rise to major differences in the dynamical tides for Saturn-like or Neptune-like planets compared to Jupiter. However, above-mentioned question is beyond the scope of this study, and we will explore this issue in future.

The scenario we propose in this work is an exotic formation mechanism for S-type planets in tight binaries ($a_{B}=0.5 \sim 3$ au). An interesting question is whether there is a limit for the separation of a binary below which in situ formation of S-type planets is impossible. Five S-type sub-Earth-sized planets have been found in Kepler-444A(BC) system \citep{Dupuy2016}. Although it has a larger $a_{B}\sim 37$ au, the periastron distance is only $\sim 5$ au ($e_B \sim 0.86$). Exotic formation mechanism such as capture seems unlikely to form a multi-planet system. Due to the close pericenter passage of 5 au, the disk of Kepler-444A will be truncated to 1-2 au. Although there are many challenges for forming such a planetary system, in situ formation can be a
possible scenario \citep{Dupuy2016}. Therefore, future planet-hunting in the binaries with a separation less than 5 au may provide vital clues to in situ or exotic formation to understand their origin.

In summary, planet formation in close binaries is a fascinating and challenging subject. As mentioned previously, searching for S-type planets in eclipsing
binaries arouses great scientific interest for the forthcoming space missions such as PLATO. Our work provides some suggestions for selecting target binaries in search for S-type planets in the near future. Their presence or absence will deepen our understanding
of planet formation.

\section*{Acknowledgments}

We thank the referee for constructive comments and suggestions. We also thank Adam P. Sutherland for helpful discussions. This work is financially supported by National Natural Science Foundation of China (Grants No. 11773081, 11573018, 11661161013), CAS Interdisciplinary Innovation Team, and the Foundation of Minor Planets of Purple Mountain Observatory. Gong Y.-X. also acknowledges the support from Shandong Provincial Natural Science Foundation, China (ZR2014JL004).


\clearpage
\appendix

\section{Code tests}

We performed additional simulations to examine whether the modified codes can accurately identify pericentre passages and exert the tide effect at each pericentre passage. In order to shorten the integration time and make the dynamical tide operate immediately, we adopt the following initial conditions. We assume that the planet has entered the stable region of the primary star in the beginning. The planet is on a very high eccentric orbit with respect to the primary ($e_{p}=0.95$, $a_{p}=0.2$). The parameters of the binary are $M_{1}=M_{\odot}$, $M_{2}=0.1 M_{\odot}$, $a_{B}=1$ au, $e_{B}=0$, respectively. The following method is used to identify that the tidal effect is added at pericentre passages. We output and record the coordinates of the planet ($\mathbf{X}_{tide}$) when the tide piece of codes is performed. Top panel of Figure A1 shows the orbital evolution of the planet. All coordinates of the planet are with respect to the primary. The red plus sign denotes the coordinates $\mathbf{X}_{tide}$ we recorded. Figure A1 clearly shows the tide is added at pericentre passages. To show the details, the time varying distance between the captured planet and the primary is plotted in the bottom panel of Figure A1 (blue line). The red line in the bottom panel is the orbital energy per unit mass of the planet over time. The Figure clearly shows the energy jump at each pericentre passage. Above tests have verified that the energy kicks caused by the dynamical tides are applied exactly in our codes.

\begin{figure}
\includegraphics[scale=0.54]{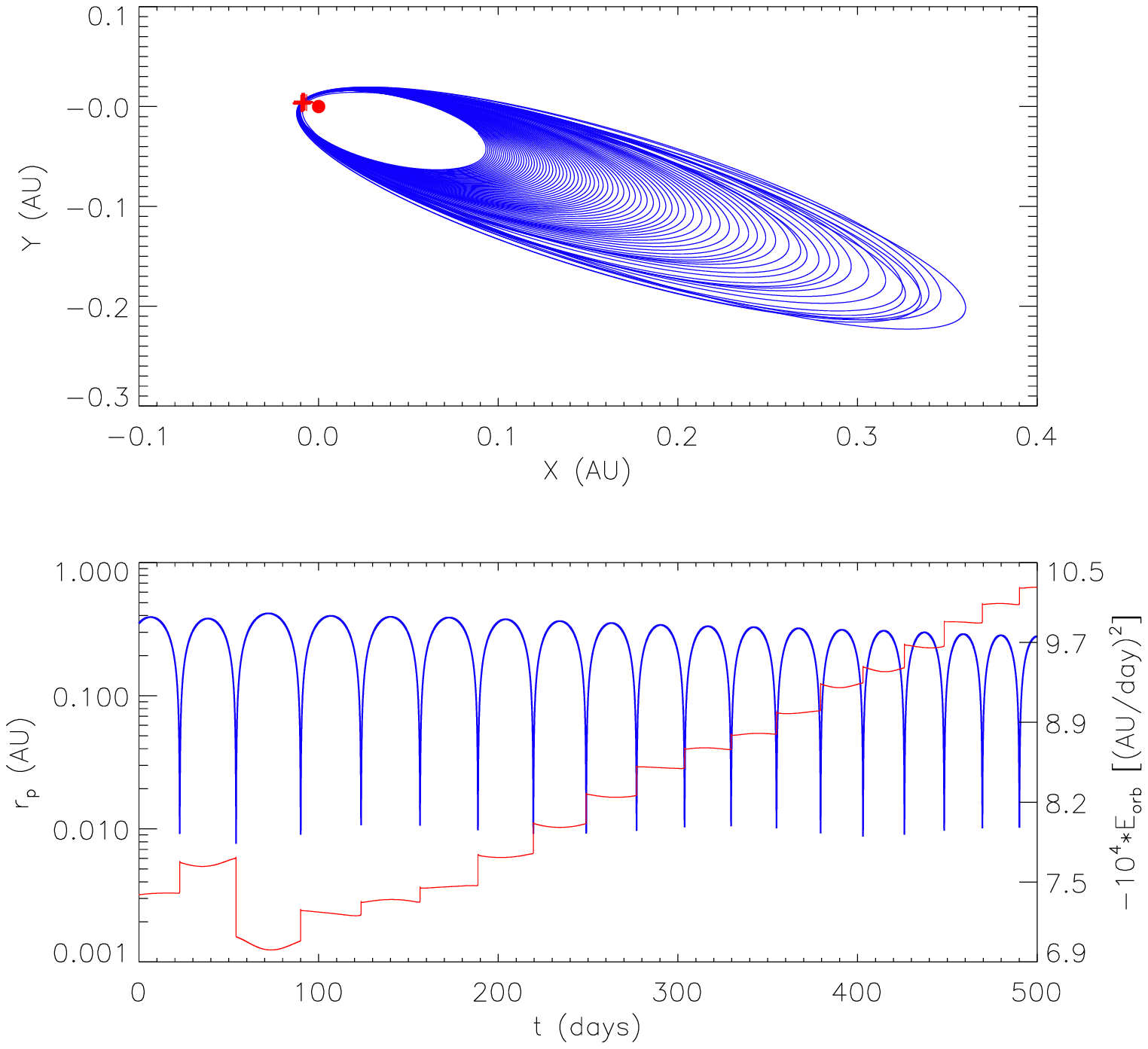}
  \centering
  \caption{Examples of pericentre passages. \textit{Top panel}: the orbital evolution of the planet. All coordinates of the planet are with respect to the primary star. The red plus sign represents the coordinates $\mathbf{X}_{tide}$ we recorded. The red dot indicates the primary star. \textit{Bottom panel}: a snapshot of the orbital evolution. The blue line shows time varying distance between the captured planet and the primary $r_{p}$. The red line is the orbital energy per unit mass of the planet (multiply by $-10^{4}$) over time. The figure clearly shows the energy kicks at pericentre passages.}
 \label{fig7}
\end{figure}


\label{lastpage}

\end{document}